\documentclass[preprint,prb,
superscriptaddress,
amsmath,amssymb,floatfix]{revtex4-1}

\usepackage{graphicx}
\usepackage{dcolumn}
\usepackage{bm}
\usepackage{ulem}

\usepackage{amssymb}
\usepackage{epstopdf}
\usepackage{textcomp}
\usepackage{enumerate}
\usepackage{epsfig,amsmath}
\usepackage{subfigure}
\usepackage[colorlinks]{hyperref}
\usepackage[usenames,dvipsnames]{xcolor}
\usepackage{float}
\usepackage{wasysym}
\usepackage{pifont}
\usepackage{tikz}
\usepackage{array}
\usepackage{colortbl}

\definecolor{Gray}{gray}{0.89}
\definecolor{plotGreen}{rgb}{0.667, 1, 0.333}
\definecolor{plotOrange}{rgb}{0.848, 0.324, 0.098}
\definecolor{plotBlue}{rgb}{0, 0.445, 0.738}
\usepackage{siunitx}

\usepackage{MnSymbol}
\usepackage{tabularx}

\begin{document}


\title{Suppression of $1/f$ noise in graphene due to non-scalar mobility fluctuations induced by impurity motion
}

\author{Masahiro Kamada}
\affiliation{Low Temperature Laboratory, Department of Applied Physics, Aalto University School of Science, P.O. Box 15100, 00076 Aalto, Finland}

\author{Weijun Zeng}
\affiliation{Low Temperature Laboratory, Department of Applied Physics, Aalto University School of Science, P.O. Box 15100, 00076 Aalto, Finland}
\affiliation{QTF Centre of Excellence, Department of Applied Physics, Aalto University 
00076 Aalto, Finland}

\author{Antti Laitinen}
\affiliation{Low Temperature Laboratory, Department of Applied Physics, Aalto University School of Science, P.O. Box 15100, 00076 Aalto, Finland}

\author{Jayanta Sarkar}
\affiliation{Low Temperature Laboratory, Department of Applied Physics, Aalto University School of Science, P.O. Box 15100, 00076 Aalto, Finland}

\author{Sheng-Shiuan Yeh}
\affiliation{International College of Semiconductor Technology, National Yang Ming Chiao Tung University, Hsinchu City, Taiwan}

\author{Kirsi Tappura}
\affiliation{Microelectronics and quantum technology, VTT Technical Research Centre of Finland Ltd., 
P.O. Box 1000, 02044 VTT, Finland}
\affiliation{
Microelectronics and quantum technology, VTT Technical Research Centre of Finland Ltd., 
QTF Centre of Excellence, P.O. Box 1000, 02044 VTT, Finland}

\author{Heikki Sepp\"a}
\affiliation{Microelectronics and quantum technology, VTT Technical Research Centre of Finland Ltd., P.O. Box 1000, 02044 VTT, Finland}

\author{Pertti Hakonen}
\affiliation{Low Temperature Laboratory, Department of Applied Physics, Aalto University School of Science, P.O. Box 15100, 00076 Aalto, Finland}
\affiliation{QTF Centre of Excellence, Department of Applied Physics, Aalto University 
00076 Aalto, Finland}
\email{pertti.hakonen@aalto.fi}

\date{\today}

\begin{abstract}
Low frequency resistance variations due to mobility fluctuations is one of the key factors of $1/f$ noise in metallic conductors. According to theory, such noise in a two-dimensional (2D) device can be suppressed to zero at small magnetic fields, implying important technological benefits for low noise 2D devices. In this work, we provide direct evidence of anisotropic mobility fluctuations by demonstrating a strong field-induced suppression of noise in a high-mobility graphene Corbino disk, even though the device displays only a tiny amount of $1/f$ noise inherently.
The suppression of the $1/f$ noise depends on charge density, showing less non-uniform mobility fluctuations away from the Dirac point with charge puddles. We model our results using a new approach based on impurity clustering dynamics and find our results consistent with the $1/f$ noise induced by scattering of carriers on mobile impurities forming clusters.

\end{abstract}

\maketitle

Modeling of $1 / f$ noise is a challenging task that has been investigated intensively since the invention of semiconducting transistors in the late 1940'ies. Typically, models based on a collection of two-level systems (TLS) or trap states are employed \cite{Kogan2008,Grasser2020}. Using a large collection of such states with broadly distributed parameters, wide band $1/f$ noise can be generated. In particular, analysis of low frequency noise in terms of charge traps in transport channels in field effect transistors has been very successful \cite{Grasser2020}.

According to the commonly accepted view, the $1/f$ noise in metallic conductors is determined entirely by fluctuations in charge carrier mobility \cite{Hooge1981a}. Several models have been put forward to elucidate the mobility noise. Apart from the models based on localized states and fluctuating scattering cross sections \cite{Hooge1994}, models based on modification of electron interference by mobile defects have been quite successful in accounting for many experimental observations \cite{Feng1986,Pelz1987}. Furthermore, noise due to agglomeration of impurities have been investigated using master equation \cite{Ruseckas2011} and lattice gas \cite{Jensen1990} type of approaches, and noise close to $1/f$ spectrum has been obtained.

In the present experimental work, our aim is to investigate the generic nature of $1/f$ noise in high-quality graphene. We will demonstrate that major part of the noise originates from mobility fluctuations, the effect of which depends on the magnetic field. Mobility fluctuations can be modeled by moving impurities or moving defects, which leads to universal characteristics due to impurity dynamics that we compare with our results. For example, the dynamics of mobile impurities may lead to genuine long-time correlations which may dominate the low-frequency noise under favorable conditions instead of a combination of random single fluctuators \cite{Ralls1988}.
Long-time memory effects are naturally offered by the nearly endless number of possible reconfigurations among a collection of mobile impurities.
The  reshaping of impurity clusters via infrequent hopping events across large energy barriers leads to long-term, non-exponential correlations, which yields $1/f$ noise over large frequency spans.

Recent experiments imply that the origin of $1/f$ noise in graphene is complex, in particular near the charge neutrality point (Dirac point) \cite{Balandin2013a,Karnatak2017}. The noise is argued to arise from an interplay of charge traps, atomic defects, short and long range scattering, as well as charge puddles. Various models have been proposed, and qualitative agreement with the data has been reached  \cite{Heller2010,Pal2011b,Zhang2011a,Kaverzin2012,Kumar2015a,Arnold2016,Karnatak2016}. In many graphene devices, even correlations between charge traps and mobility noise have been found \cite{pellegrini2013,Lu2014}, which is also common in regular metallic and semiconducting devices \cite{Dutta1981,Hooge1994,Kogan2008}. In our work, suspended clean graphene removes many of these noise sources and the fundamental noise elements can be addressed in pure form.

Corbino geometry is unique for electrical transport as the magnetoresistance includes only the diagonal conductivity component $\sigma_{xx}$. This feature means that, if there are isotropic mobility fluctuations (fully correlated in two orthogonal directions) causing $1/f$ noise, this noise component will be suppressed to zero at $\mu_0 B=1$ \cite{Orlov1992}; here $\mu_0$ denotes the mobility at $B=0$. This behavior has unsuccessfully been searched for both in two-dimensional electron gas \cite{Levinshtein1983,Song1985,Song1988,Orlov1990} as well as in graphene \cite{Rumyantsev2013}. Our experimental results display a clear suppression of noise as a function of $B$, with a minimum around $\mu_0 B \simeq 1$. This is direct proof that a large part of the $1/f$ noise originates from mobility fluctuations in clean graphene.

\section{1/f Noise}

In metallic materials, the  power spectral density of $1/f$ fluctuations is often assigned to the random impurity scattering due to the mobile
impurities, defects, or vacancies \cite{Fleetwood2015}. Accordingly, the spectral density of $1/f$ noise $S_{m}(f)$ is related to fluctuations of the conductivity $\sigma_{m}$  governed by mobile impurities,
\begin{equation}
S_{m}(f )
=\frac{\left\langle \Delta \sigma _{m}^{2}\right\rangle }{\sigma _{m}^{2}}
=\frac{\left\langle \Delta \mu _{m}^{2}\right\rangle }{\mu _{m}^{2}}
=\mathbf{s}_{m}\frac{1}{f},
\end{equation}%
where $\sigma_{m}=e\mu _{m}n$ with $\mu _{m}$ limited by scattering of the mobile impurities alone, $e$ is the electron charge, and $\mathbf{s}_m=\mathrm{const.}$  (when the number of impurities is fixed), i.e. the noise does not scale with the density of charge carriers $n$. Thus, parallel to works of Refs. \onlinecite{Fleetwood1985,Pelz1987}, our starting point differs from that of Hooge \cite{Hooge1972,Hooge1981a} who argued that the $1/f$ noise due to mobility fluctuations varies inversely with the total number of charge carriers $N_e$: $\mathbf{s}_m=\mathrm{constant}/N_e$. 
The conductivity $\sigma_{m}$ describes, in general,
transport associated with scattering from a disordered and
time-dependent part of a system. Constant $\mathbf{s}_{m}$\
depends on temperature (diffusion constant), number of impurities,
interaction between them, volume or the area of the component as well as the
details related to the scattering process. Our hypothesis is that in a specific sample under fixed external conditions, $\mathbf{s}_{m}$ remains unchanged, for example even though the sample resistance is changed significantly by enhancing the carrier number by gating. If scattering is anisotropic, then correlations among $1/f$ noise contributions may arise, which yield specific characteristics for the noise when external conditions are changed (see below). Note that a magnetic field $ B$ changes the sample, because the length $L$ of the traversed carrier path becomes longer. Consequently, in the incoherent transport case, the noise due to mobile impurities becomes suppressed as $\mathbf{s}_{m} \propto 1/L \propto 1/R(B)$.  

Conductance in graphene is influenced by immobile impurity
scattering, both due to short-ranged and Coulomb scatterers, as well as randomly moving mobile impurities, with the scattering rates proportional to inverse mobilities $1/\mu _{s}$, $1/\mu _{C}$, and $1/\mu_{m}$, respectively. For simplicity, we neglect here the electron-phonon scattering which, however, may govern the inelastic scattering length that is important in electronic interference effects. The graphene conductivity is then given by $\frac{1}{\sigma _{g}}=\frac{1}{ne}\left( \frac{1}{\mu _{C}}+\frac{1%
}{\mu _{s}}+\frac{1}{\mu _{m}}\right)$
according to the Mathiessen rule. Consequently, the conductance fluctuation of graphene can be written as%
\begin{equation}\label{Rgr}
S_{g}(f )=\frac{\left\langle \Delta \sigma_{g}^{2}\right\rangle }{\sigma_{g}^{2}}
=\left(\frac{\sigma_g}{\sigma_m} \right)^2\frac{\left\langle  \Delta \sigma _{m}^{2}\right\rangle }{\sigma_{m}^{2}}
=\left(\frac{\mu_g}{\mu_m} \right)^2\frac{\left\langle  \Delta \mu_{m}^{2}\right\rangle }{\mu_{m}^{2}}
=\left(\frac{\mu_g}{\mu_m} \right)^2 \mathbf{s}_{m}\frac{1}{f},
\end{equation}%
%
where $\frac{1}{\sigma _{g}}=\frac{1}{ne} \frac{1}{\mu _{g}}$.
The noise of a graphene device thus depends on the electron mobility due to mobile impurity scattering as well as their relative significance in total conductivity given in terms of total graphene mobility $\mu_g$; if this ratio is modified by some physical process, then the magnitude of the $1/f$ noise changes. For example, if mobile impurities behave as short range scatterers, the mobility related with them may decrease with increasing gate voltage $V_g$ \cite{MGres2020}.

In a typical graphene sample, contact resistance starts to become important at large charge densities at which $\sigma_g>> G_0$; $G_0=2e^2/h$ denotes the conductance quantum. Consequently, $1/f$ noise from contacts has to be taken into account, particularly at large $V_g$. We model the contact resistance by two parallel trasport pathways, one for electrons and one for holes, with conductance $G_{ce}$ and $G_{ch}$, respectively. We assume that $G_c(n)=G_{ce}(n)+G_{ch}(n)=G_{c}=\mathrm{const}$. The electrons and holes are treated separately so that their noise is added incoherently, which yields  
\begin{equation} \label{Scon}
S_{c}(f)=\frac{\left\langle \Delta G_{c}^{2}\right\rangle }{%
G_{c}^{2}}=\frac{\left\langle \Delta R_{c}^{2}\right\rangle }{%
R_{c}^{2}}=\frac{n_{h}^{2}+n_{e}^{2}%
}{\left( n_{h}+n_{e}\right) ^{2}}\frac{\mathbf{s}_{c}}{f},
\end{equation}%
where $n_{e}$ and $n_{h}$ specify the number density of electrons and holes and we have assigned equally large noise constant $\mathbf{s}_{c}$ to both carrier species at the two contacts:
$
S_{ch}(f) = S_{ce}(f )= \frac{\left\langle \Delta R_{ch}^{2}\right\rangle }{R_{ch}^{2}}
= \frac{\left\langle \Delta R_{ce}^{2}\right\rangle}{R_{ce}^{2}}
=\frac{\mathbf{s}_{c}}{f}$.
Here $R_{c}=1/G_{c}$ denotes the contact resistance, while its components $R_{ce}=1/G_{ce}$ and $R_{ch}=1/G_{ch}$ specify the contact resistance for electrons and holes, respectively. 
The density of charge carriers as function of the gate voltage can be estimated as $n =C_{g}\left[ \sqrt{V_{g}^{2}+V_{d}^{2}}\right]/e$, where the electron-hole crossover voltage scale $V_d  \simeq 1.4$ V is close to the residual charge range determined from the measured $G(V_g)$. This $V_d$ scale determines the coexistence of electrons and holes according to 
$n_{e(h)} = \frac{1}{2}C_{g}\left[ \sqrt{V_{g}^{2}+V_{d}^{2}}+(-)V_{g}\right]/e$, which fulfills $n=n_e + n_h$. 
Note
that the constant $\mathbf{s}_{c}$ has a slightly different
role than the constant $\mathbf{s}_{m}$ since we do not divide the
contact resistance into parts according the type of the scattering
mechanisms. This is because the contact resistance remains a constant in the range of interest. 
Near the Dirac point the equation can be approximated as $
S_{c}(f)\approx \left( 1-2\frac{n_{e}n_{h}}{n^{2}}\right) \mathbf{s}_{c}\frac{1}{f}$
which shows that the noise reaches a minimum at the Dirac point. Without invoking Hooge's law (cf. Refs. \onlinecite{Xu2010,Zhang2011a}), the above equation provides a simple explanation for the noise dip as a consequence of incoherent transport of electrons and holes through a noisy contact resistance.

On the whole, $1/f$ noise in our model is obtained by combing contributions from Eqs. \ref{Rgr} and \ref{Scon} incoherently. Neglecting the variation of $\mu_m$ (though, see below), the noise is described using two fit parameters, $\mathbf{s}_m$ and $\mathbf{s}_c$, while the rest of the parameters are determined from the conductance $G(V_g)$. Besides these two parameters, we employ a correlation factor $\chi$ in order to account for magnetic field dependence of the noise, which clearly indicates the presence of mobility fluctuations: $\chi$ describes the correlations between two orthogonal mobility fluctuation components. Using these three parameters we are able to account quite well for our measured noise results covering the parameter range of $|n| < 4 \times 10^{14}$ m$^{-2}$ and $0 \leqq B< 0.15$ T.

\section{\label{sec:setting}Experimental setting}

\begin{figure}[t!] 
	\centering
	\includegraphics[width=.85\linewidth]{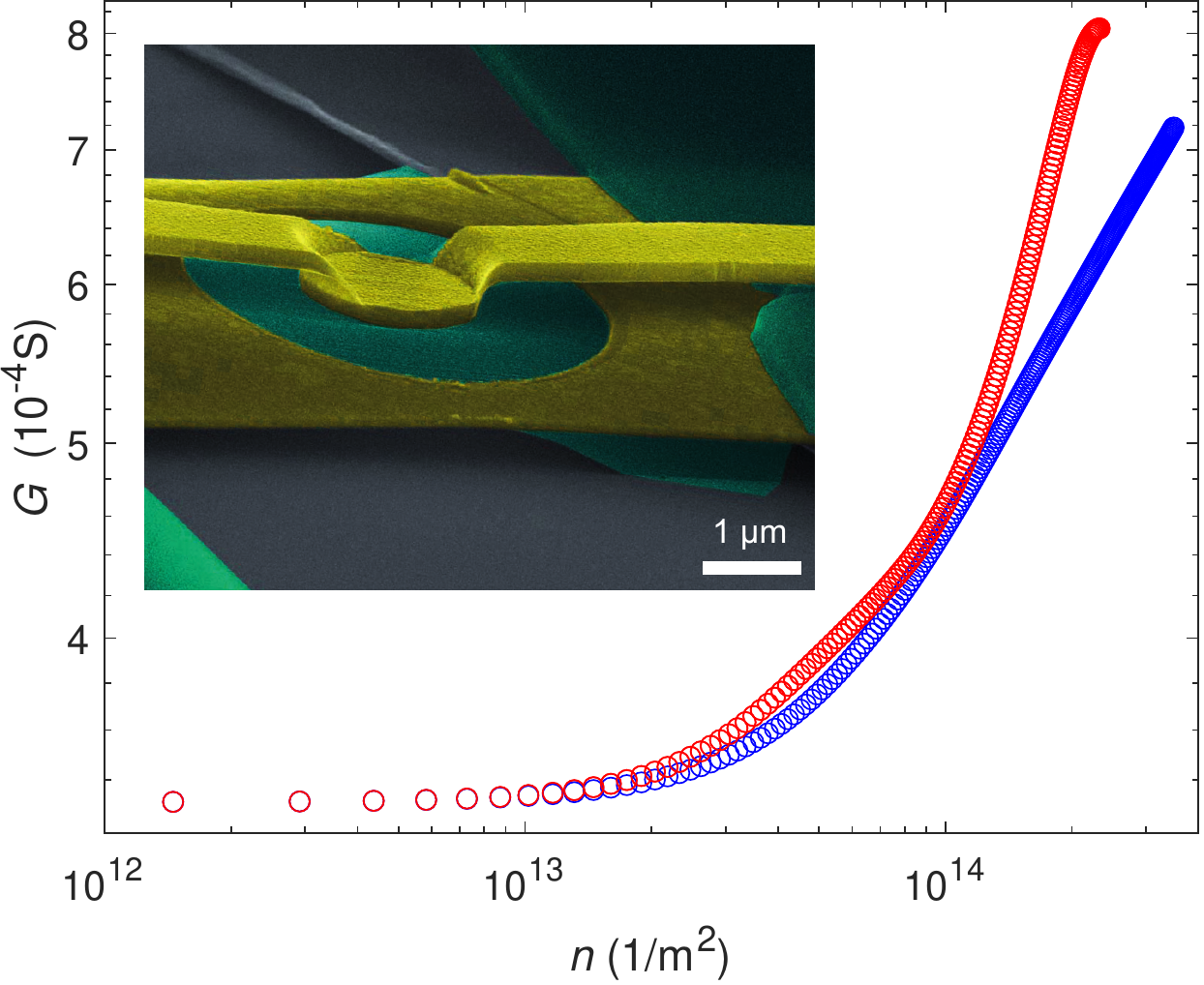}
	\caption{ 
	Zero-bias conductance $G$ vs. charge carrier density $n$ measured at $T=4.2$ K for $V_g > 0$ ({\textcolor{red}{$\circ$}}) and $V_g < 0$ ({\textcolor{blue}{$\circ$}}). The inset displays a scanning electron microscope image of our suspended graphene Corbino disk (green part in the center) with \SI{4.5}{\micro\metre} diameter outer Cr/Au contact and \SI{1.8}{\micro\metre} diameter inner Cr/Au contact. Scale bar for \SI{1}{\micro\metre} is given in the image.
	\label{sample}}
\end{figure}

A scanning electron microscope picture of our graphene Corbino sample with Cr/Au electrodes is illustrated in the inset of Fig. \ref{sample}: the size of the disk is \SI{4.5}{\micro\metre} in outer diameter and \SI{1.8}{\micro\metre} in inner diameter. 
The gate voltage dependence of the conductance $G(V_g)$ of our sample at $B=0$ (see Fig. \ref{sample}) yielded for
the residual charge carrier density $n_0 = 0.5\textup{--}1 \times 10^{14}$ m$^{-2}$. Details of the sample and its fabrication are discussed in the Methods section.

Our Corbino device displays strong classical, geometric magnetoresistance given by $R_g(B)=R_g\left(1+(\mu_0 B)^2\right)+R_c$ where $R_g$ and $\mu_0$ denote the resistance and mobility of graphene at zero field, respectively \cite{MGres2020}. Following the analysis of Ref. \onlinecite{MGres2020}, we determine the contact resistance using the well known diffusive transport modeling for graphene with definite gate dependence for impurity scattering \cite{Hwang2007,MGres2020} and obtain $R_c \simeq 800$ $\Omega$, almost independent of gate voltage in the range of gate voltages $V_g = \textup{--}4 \dots \textup{+}4$ V. Also, we set $R_{ce}  =R_{ch} = R_{c}$.
The slope of the resistance change  $\Delta R_g(B)/R_g$ vs $B^2$, where $\Delta R_g(B)=R_g(B)-R_g$, indicates that $\mu_0\simeq 1\textup{--}2 \times 10^5$ cm$^2$/Vs is approximately constant. The strong $B^2$ magnetoresistance can also be viewed as a good indicator of the high quality of our sample, as the magnetoresistance in disordered graphene would display  more involved field dependence \cite{Ando2002,Sachdev2008,Weber2012, Alekseev2013}.
Since the role of transverse conductivity components vanishes in the Corbino geometry, the mobility is equivalent to that obtained in Hall measurements. Since the ratio of azimuthal ($\hat{\varphi}$) to radial ($\hat{r}$) direction currents is given by $i_{\varphi}/i_r = \mu_0 B$, the current path through the sample is substantially lengthened with growing $\mu_0 B$.

\section{Results}

\begin{figure}[t!]
	\centering
    \includegraphics[width=.40\linewidth]{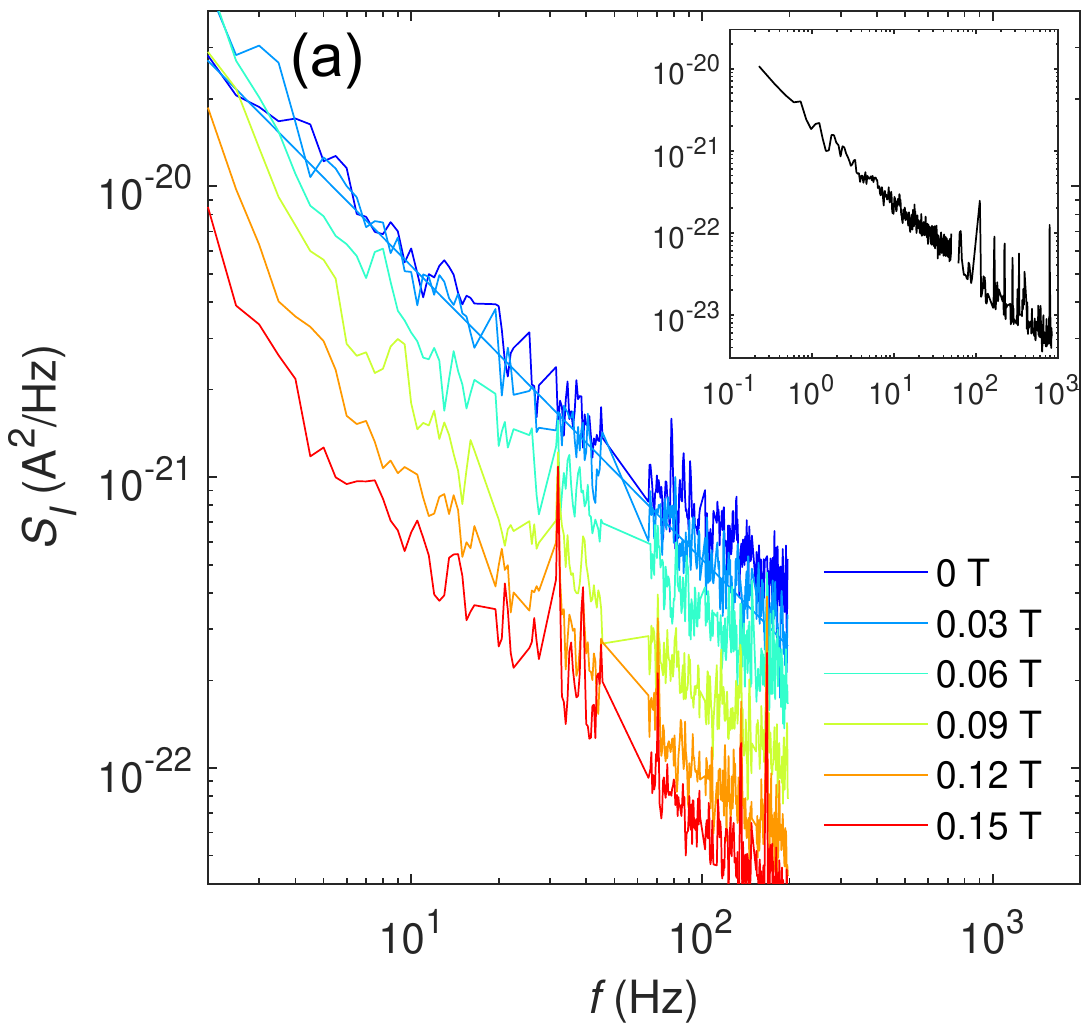}
	\includegraphics[width=.47\linewidth]{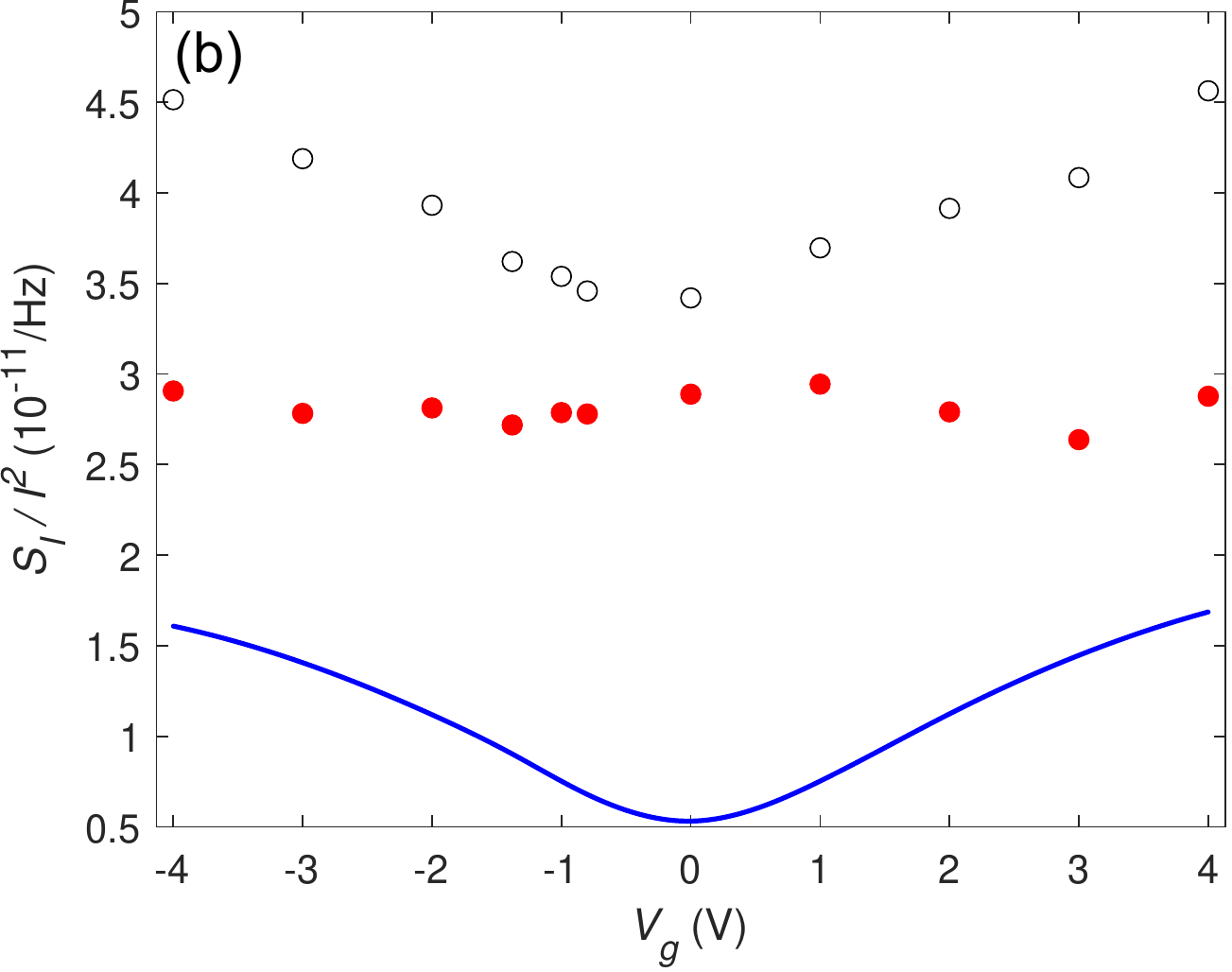}

	\caption{a)
	Noise power spectral density $S_I$ vs. frequency $f$ at a few fields between 0 and 0.15 T measured at $V_g =$ 10 V and at currents between 3.78 and \SI{10.7}{\micro\ampere}; the current is reduced due to geometric $B^2$ magnetoresistance. For the data at 0.03 T ($I=10.2$ \SI{}{\micro\ampere}), the result of $1/f^{\beta}$ noise fitting with $\beta = 1.01$ is also shown. The inset depicts $1/f$ noise scanned over three decades in frequency at $I=9.3$ \SI{}{\micro\ampere}.
	b) Measured (at $T=27$ K), scaled $1/f$ noise spectral density $S_I/I^2$ illustrated at 10 Hz as a function of gate voltage $V_g$ ({\textcolor{black}{$\circ$}}); $|n| < 4 \times 10^{14}$ m$^{-2}$. The filled circles ({\textcolor{red}{$\bullet$}}) denote the noise contribution due to graphene resistance fluctuations $\frac{R_{g}^2}{R_{\Sigma}^2}S_I^g/I^2 = \frac{\delta R_{g}^2}{R_{\Sigma}^2}$ while the solid (blue) curve  displays the current noise contribution from contact resistance $\frac{R_{c}^2}{R_{\Sigma}^2}S_I^c/I^2 = \frac{\delta R_{c}^2}{R_{\Sigma}^2}$.
	\label{NoiseB0} }
\end{figure}

Fig. \ref{NoiseB0}a displays measured noise power spectral density $S_I$ vs. frequency $f$ at a few fields between 0 and 0.15 T at $V_g =$ 10 V using currents around \SIrange[range-units = single,range-phrase=\textup{--}]{5}{10}{\micro\ampere}.
All the data are very close to the pure $1/f$ form: for the data at 0.03 T and $ I=10.2$ \SI{}{\micro\ampere}, for example, fitting of a free exponent $\beta$ to $1/f^{\beta}$ yields $\beta = 1.01$. The inset displays a wider frequency scan of the noise at $I=9.3$ \SI{}{\micro\ampere}: clear  $1/f$ noise is present over three decades in frequency. 
Our data also fulfill the basic properties of $1/f$ noise spectra as a function of bias current $S_I(I) = \mathbf{s} \frac{ I^{\gamma}}{ f}$, where $\gamma \sim 2$. The noise magnitude $\mathbf{s}=f \times S_I(I)/I^2 \simeq 2\textup{--}5 \times 10^{-10}$ is approximately a constant, which is decomposed into $\mathbf{s}_m$ and $\mathbf{s}_c$ in our analysis. Compared with other low-noise graphene devices, the noise of our sample is on par with the lowest achieved results \cite{Kumar2015a,Kamada2021}.

The $1/f$ noise changed irregularly with the increase of magnetic field at low temperatures, in particular at 4 K and below. This is assigned to the role of conductance fluctuations in our sample. Their significance becomes reduced with increasing temperature as the thermal diffusion length $L_T=\sqrt{\frac{\hbar D}{k_B T}}$ decreases from $400$ to about $ 100$ nm when $T$ is increased from 4 to 40 K. Consequently, we selected an operation temperature of $T=27$ K, at which the strength of individual conductance fluctuation features was sufficiently weakened, and the magnetic field dependence could be analyzed better in terms of diffusive transport models.

Charge density dependence of the scaled $1/f$ noise spectral density, $\frac{R_{g}^2}{R_{\Sigma}^2}S_I^g/I^2 = \frac{\delta R_{g}^2}{R_{\Sigma}^2}$ and $\frac{R_{c}^2}{R_{\Sigma}^2}S_I^c/I^2 = \frac{\delta R_{c}^2}{R_{\Sigma}^2}$, where $R_{\Sigma}=R_{g}+R_{c}$, is depicted in Fig. \ref{NoiseB0}b for the graphene part and the contacts, respectively. The value for contact noise $S_I^c/I^2= 3.17 \times 10^{-10}/f$ was determined from data measured at $V_g=\textup{--}70$ V in the unipolar regime where no $pn$ interfaces exist and $\frac{R_{c}^2}{R_{\Sigma}^2} \sim 1$. The data points specified at $f=10$ Hz were taken from the fits of $1/f$ form to the measured spectra. The open circles denote the experimental data at 27 K, while the filled symbols and the blue curve indicate the separation between the noise contributions from the graphene itself and the contacts, respectively. Clearly, graphene contribution dominates the measured  $1/f$ noise at the Dirac point, but at $|n| \sim 4 \times 10^{14}$ m$^{-2}$ the graphene contacts account for $\sim 1/3$ of the noise. As the graphene resistance increases with $B$, the contact contribution in the coupled noise becomes even less significant in a magnetic field. The dip structure in the noise near the Dirac point is due to the substantial magnitude of $S_I^c/I^2$ compared with the graphene noise, and it originates from the presence of both electron and hole carriers and incoherent addition of their contact noise contributions.

\begin{figure}[bt!]
	\centering
	\includegraphics[width=.47\linewidth]{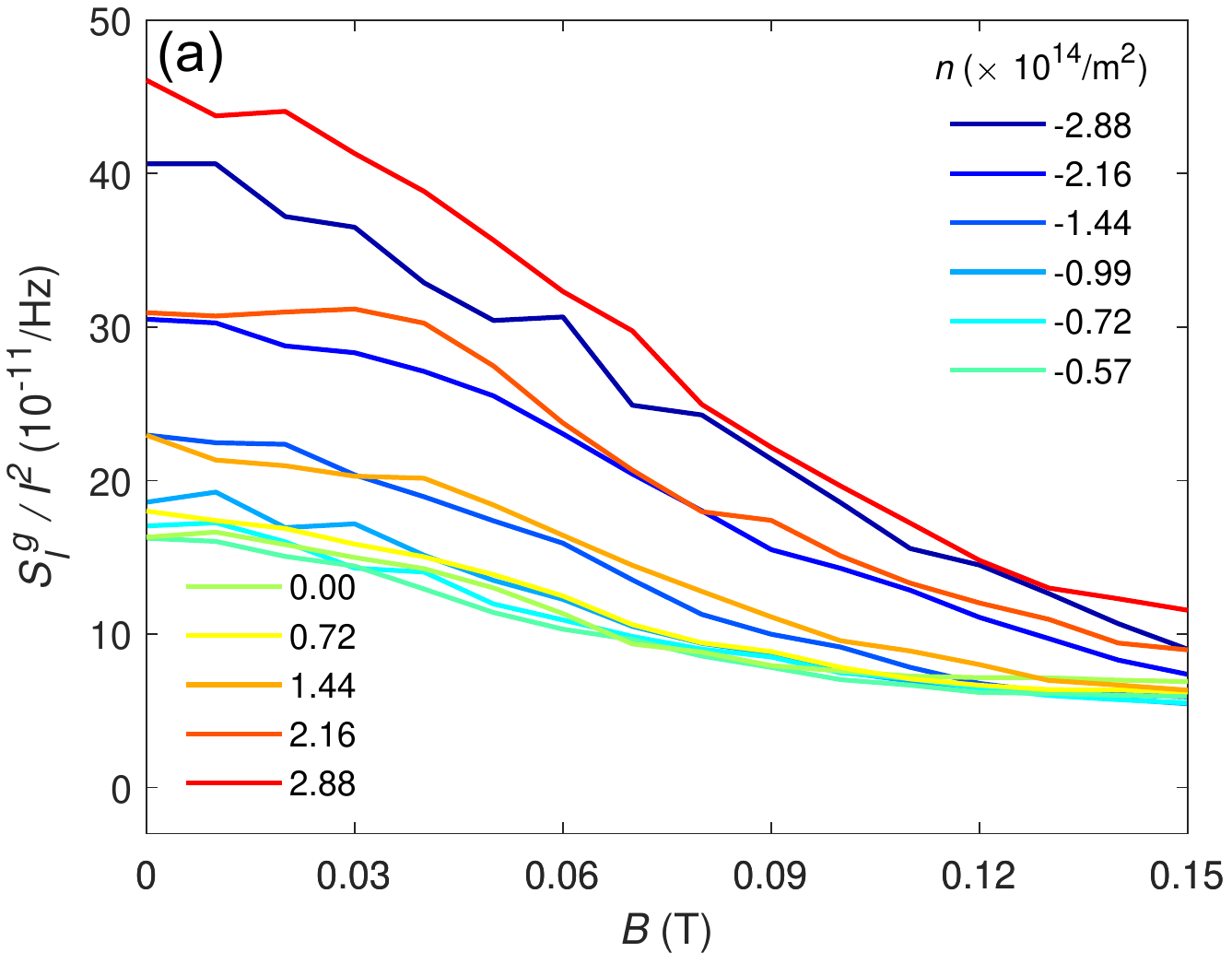}
	\includegraphics[width=.47\linewidth]{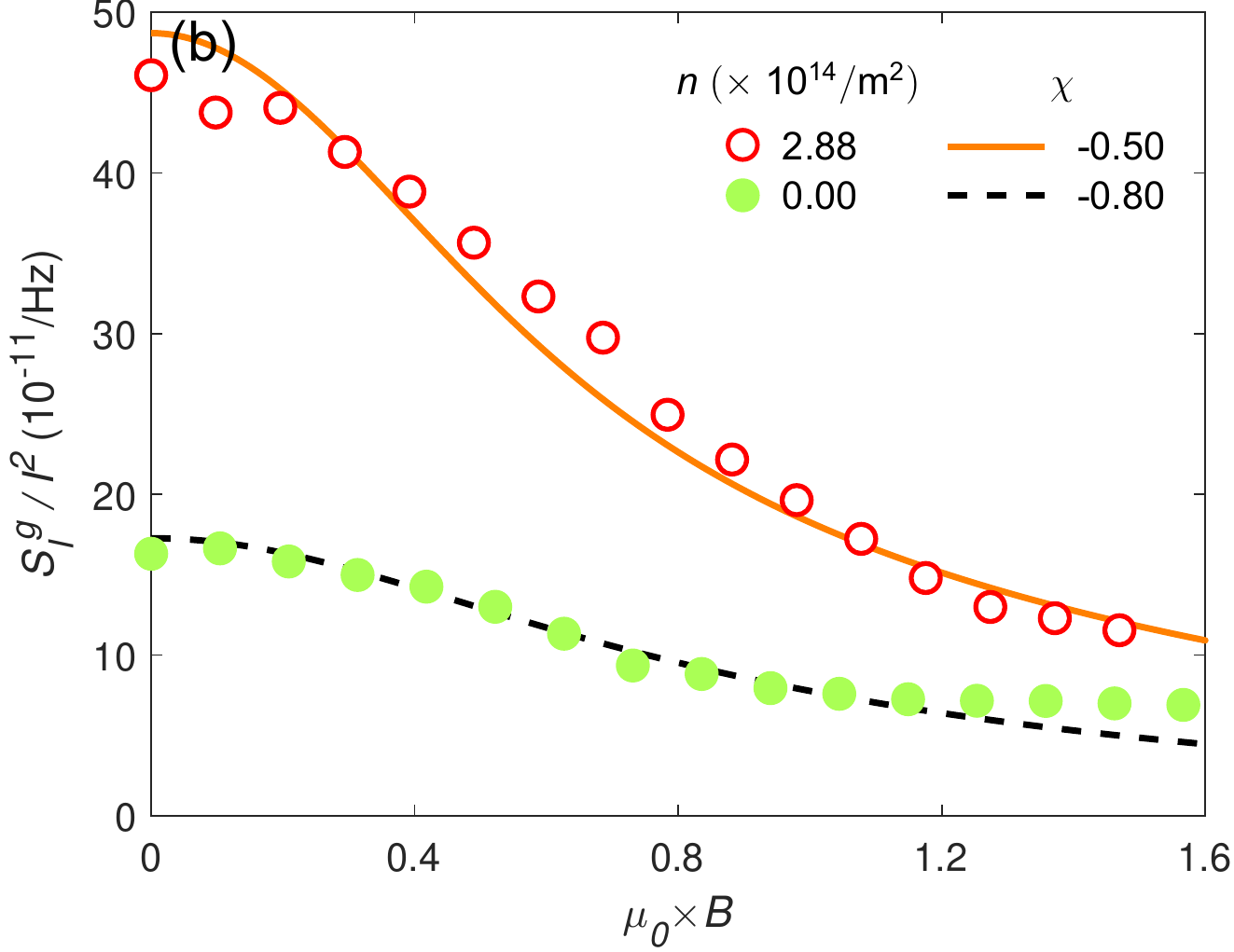}
	\caption{a) Scaled spectral density of graphene noise $S_I^g/I^2$ at 10 Hz vs. magnetic field $B$ at various nominal $V_g$-induced charge densities indicated in the two insets.  
	b) Data at Dirac point ({\color{plotGreen}$\bullet$}) and at $n = 2.88 \times 10^{14}$ m$^{-2}$ ({\color{red}$\circ$}) plotted as a function of product $\mu_0 \times B$. The dashed and solid curves illustrate theoretical behavior of Eq. (4) using correlation coefficient of $\chi=-0.8$ and $\chi=-0.5$, respectively.  Measurement temperature was $T = 27$ K.
	\label{S_I2_mu_vs_B}}
\end{figure}

At small magnetic fields, $B<0.15$ T, the resistance grows as $B^2$ and, simultaneously, the apparent Dirac point shifts slightly higher in $V_g$ \cite{MGres2020}. 
The magnetic field dependence of the measured $1/f$ noise is illustrated in Fig. \ref{S_I2_mu_vs_B}a, which depicts scaled graphene noise power $S_I^g/I^2=  \frac{R_{\Sigma}^2}{R_{g}^2}S_I/I^2-\frac{R_{c}^2}{R_{g}^2}S_I^c/I^2$ at 10 Hz as a function of magnetic field $B$ at several charge carrier densities near the Dirac point ($ n = \textup{--}3 \dots \textup{+}3 \times 10^{14}$ m$^{-2}$). A clear reduction in noise is observed with increasing $B$ and the reduction takes place in approximately equal relative manner at all gate voltage values. The smallest noise $S_I^g/I^2$ is observed at the Dirac point, as typical, while the reduction of noise with $B$ becomes strongest at charge densities $n > \pm 1.5 \times 10^{14}$ m$^{-2}$ at which the reduction by the field amounts to $\sim 75\%$. A  leveling off of the reduction appears when approaching $B = 0.15$ T, but no clear upturn is found in the investigated $B<0.15$ T range. $B = 0.15$ T was selected as the upper limit in our analysis because there was already a deviation visible from the $B^2$ dependence in the magnetoresistance $\Delta R/R$.

For Corbino devices with isotropic (scalar) mobility fluctuations $\delta \mu_0$, the current fluctuations should vanish at a sweet spot having $\mu_0 B = 1$. In practice, the sweet spot does not reduce the noise down to zero, but non-idealities/other noise sources will limit the reduction \cite{Levinshtein1983,Song1985,Song1988,Orlov1990,Rumyantsev2013}. On the other hand, partial correlations between mobility fluctuations in radial $\delta \mu_r$ and azimuthal $\delta \mu_{\varphi}$ direction may account for the observed noise reduction.
%
If we simply calculate the total resistance fluctuations as a linear sum of
local fluctuations of resistivity we obtain for total resistance fluctuations 
\begin{equation}\label{suppr}
S_g(f)= \mathbf{s}_{m}\frac{1}{f} \left( \frac{\mu _{g} }{\mu _{m}}\right) ^{2} \frac{\left(
1+\left( \mu_0 B\right) ^{4}-2\chi\left( \mu_0 B\right) ^{2}\right)}{\left( 1+\left( \mu_0
	B\right) ^{2}\right)^2},
\end{equation}%
where 
we characterize the combined fluctuations of $\delta \mu _{\varphi }$ and $\delta \mu_{r}$ by correlation coefficient  $\chi$ according to $\left\langle \delta \mu _{\varphi }\delta \mu _{r }\right\rangle = \chi \left\langle \delta \mu _{r}^{2}\right\rangle= \chi \left\langle \delta \mu _{\varphi}^{2}\right\rangle$; similar formulas can be derived starting from anisotropic scattering \cite{Orlov1992}.
When there is full positive (scalar) correlation ($\chi = 1$), the noise vanishes at $\mu_0 B = 1$.  With full negative correlation $\chi = -1$, no suppression is seen in the $1/f$ noise \cite{Levinshtein1983,Song1985,Song1988,Orlov1990,Rumyantsev2013}. By taking into account $\mathbf{s}_{m} \propto 1/{\left( 1+\left( \mu_0
B\right) ^{2}\right)}$, our results in Fig. \ref{S_I2_mu_vs_B}a yield $\chi \simeq -0.5 \pm 0.1$ away from the Dirac point. This result is in agreement with theoretical studies for changes in resistance due to reorganization of atoms \cite{MARTIN1972,Nagaev1982}$^,$\cite{Orlov1990} and also with noise correlation measurements in semimetallic Bi \cite{Black1983}. According to our own kinetic Monte Carlo simulations (see Methods), the correlation coefficient is $\chi \simeq -0.7$ and $-0.4$ for $k_B T/E_d =0.3$ and at $k_B T/E_d = 0.5$, respectively; here $E_d$ is the energy barrier for hopping between sites. The difference in these calculated values for $\chi$ can be explained by the fact that, at the lower temperature, a significant number of defects spend a considerable time at the electrodes forming elongated clusters instead of moving freely on graphene without additional restrictions on the shape of clusters as discussed in the supplementary information of Ref. \onlinecite{Kamada2021}.   

\begin{figure}[t!]
	\centering
	\includegraphics[width=.92\linewidth]{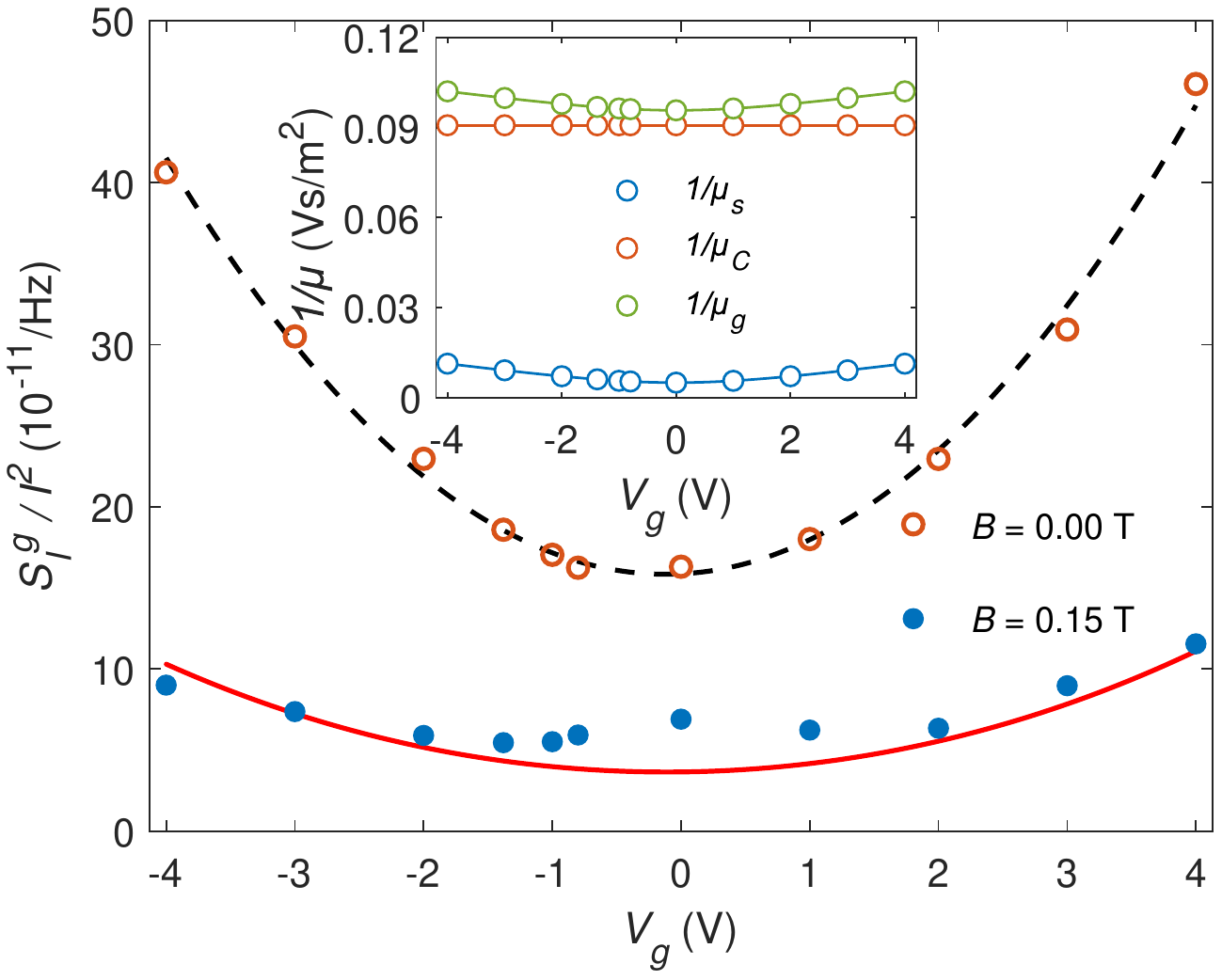}	
	\caption{Scaled graphene noise $S_I^g/I^2$ at 10 Hz versus gate voltage measured at $B=0$ ({\color{plotOrange}$\circ$}) and at $B=0.15$ T ({\color{plotBlue}$\bullet$}). The dashed parabolic curve on the  $B=0$ data reflects the expected behavior of noise due to the prefactor $\left(\frac{\mu_g}{\mu_m} \right)^2$ in Eq. \ref{Rgr}. The  solid curve overlaid on $B=0.15$ T data is calculated using Eqs.  \ref{Rgr} and \ref{suppr} with a fixed 
	$\chi=-0.5$, though slighly stronger anticorrelation is observed near the Dirac point.
The inset illustrates $V_g$ dependence of the inverse mobility $1/\mu_g$ determined at 27 K; $\mu_s$ and $\mu_C$ denote mobility limited by short-ranged and Coulomb scattering, respectively.  \label{S_I2_vs_Vg}}
\end{figure}

The gate voltage dependence of $1/f$ noise at $B=0$ and $B=0.15$ T is compared in Fig. \ref{S_I2_vs_Vg}. 
We observe that the difference between the data at $B = 0$ and $B = 0.15$ T grows monotonically when moving away from the Dirac point. 
The resistance fluctuations are increasingly suppressed from $B=0$ value with growing charge carrier density over  $|n|=  1 \dots 3 \times 10^{14}$ m$^{-2}$. This decrease can be assigned to a small change in the magnitude of correlations between $\delta \mu_r$ and $\delta \mu_{\varphi}$ from $\chi \sim -0.8 $ (at Dirac point) to $\chi \simeq -0.5$ (at $|n|= 3 \times 10^{14}$ m$^{-2}$). Possibly, the strengthening in anticorrelation between $\delta \mu _{\varphi }$ and $\delta \mu_{r}$ is related to reduced screening at small charge densities.

Graphene properties may also become modified due to external factors, such as strain induced by $V_g$ or local strains induced by adsorption of adatoms \cite{Krasheninnikov2011}. According to Ref. \onlinecite{Krasheninnikov2011}, the stress field induced by the adatom extends over till the next nearest lattice sites and it involves asymmetry in the strain distribution. Such strain leads to pseudomagnetic fields \cite{Katsnelson2012}, which act as weak uniform scattering regions. In addition, there arises scalar (gauge) fields which may increase conductivity by enhancing local charge density \cite{Kamada2021}. Hence, adsorbed mobile atoms could provide short-ranged scattering centers which cause scalar-type mobility fluctuations,
weakening negative correlations.
Detailed theoretical modeling of the actual scattering centers and the ensuing $\chi(n)$ is beyond the scope of the present work. 

According to Eq. \ref{Rgr}, graphene noise at $B=0$ varies as $\left(\frac{\mu_g}{\mu_m} \right)^2 \mathbf{s}_{m}\frac{1}{f}$, where the scattering contribution due to mobile impurities, proportional to $1/\mu_m$, may vary with gate voltage. Using diffusive transport theory \cite{Hwang2007,MGres2020}, scattering by short range impurities produces $1/\mu_s= a|n|$ while for Coulomb impurities $1/\mu_C= b$, where $a$ and $b$ are constants. Using the mobility analysis described in Ref. \onlinecite{MGres2020}, we can determine $\mu_g$, $\mu_C$ and $\mu_s$. The division between relative scattering rates in terms of components $\mu_C^{-1}$ and $\mu_s^{-1}$ at 27 K is indicated in the inset of Fig. \ref{S_I2_vs_Vg}. If we assume that mobile impurities would be Coulomb scatterers with constant $\mu_C$, $S_I^g$ would decrease with $V_g$. Hence we conclude that the majority of mobile impurities in our device must be short range scatterers due to adatoms. For short-ranged scatterers, we obtain a parabolic change of $\left(\frac{\mu_g}{\mu_m} \right)^2$ upto $300 \dots 400$\% across the measured gate voltage range (see Fig. \ref{S_I2_vs_Vg}). This change will become reduced if part of the scatterers are Coulomb impurities. In fact, the adatoms will produce both short range scattering due to strain and long range scattering due to induced charge at the impurity site. Thus, the sharp division between short range and long range contributions is not properly valid for the mobile impurities present in our system. Nevertheless, the factor $\left(\frac{\mu_g}{\mu_m} \right)^2$ agrees with the functional form of the measured change, supporting the basic assumption of $\mathbf{s}_{m}= \mathrm{const.}$ in our $1/f$ noise model. For better testing between theory and experiment, a separate means to determine the scattering contribution of the mobile impurities would be needed.

\section{Conclusions}
We have addressed the fundamental sources of $1/f$ noise in a low-noise, suspended graphene Corbino disk without strong two-level systems. 
By employing incoherent transport of electrons and holes, we could account for the contact noise by a single parameter. Also, we were able to account for the graphene noise, basically using a single parameter which is related to impurity movement and mobility fluctuations due to random agglomeration/deagglomeration of impurities. The noise decreased with increasing magnetic field, which pinpoints the nature of the noise to mobility fluctuations with intrinsic correlations between radial $\mu_r$ and azimuthal $\mu_{\varphi}$ fluctuations. We find negative orthogonal mobility correlations, which agrees with expectations for "rotating" impurities and our kinetic Monte Carlo simulations. 
With the significant noise suppression as a function of magnetic field, our work constitutes a strong direct demonstration of mobility noise in two dimensional materials. 

Further work is needed to shed light on cluster formation of defects and collective dynamics of such clusters. Recent progress in both metallic and semiconducting 2D materials facilitates good opportunities in tackling these problems. The complex, and possibly self-limiting, dynamics of impurities provides a natural explanation for the long-time memory effects needed to create genuine $1/f$ noise, distinct from the regular $1/f$ noise theories based on distributed two-level or trap states. Our results demonstrate that these collective phenomena may be addressed in very clean micron scale systems.

\section*{Acknowledgements}
We are grateful to Elisabetta Paladino, Igor Gornyi, Manohar Kumar, and Tapio Ala-Nissil\"a for fruitful discussions and to Sergey Rumyantsev for pointing Ref. \onlinecite{Levinshtein1983} to us. This work was supported by the Academy of Finland projects 314448 (BOLOSE), 310086 (LTnoise) and 312295 (CoE, Quantum Technology Finland) as well as by ERC (grant no. 670743). The research leading to these results has received funding from the European Union’s Horizon 2020 Research and Innovation Programme, under Grant Agreement no 824109. The experimental work benefited from
  the Aalto University OtaNano/LTL infrastructure.
A.L.\ is grateful to V{\"a}is{\"a}l{\"a} foundation of the Finnish Academy of Science and Letters for scholarship.

\section*{Methods}

All the suspended monolayer graphene devices employed in this work were fabricated using a technique based on lift off resist (LOR) sacrificial layer \cite{Tombros2011}. Details on our sample fabrication process can be found in Ref. \onlinecite{Kumar2018}. First, graphene was exfoliated on LOR (thickness 500 nm) using a heat-assisted exfoliation technique \cite{Huang2015} and characterized using Raman spectroscopy. Electron-beam lithography was employed to pattern the contacts ($5$ nm Cr/ $60$ nm Au) using a PMMA 50k/950k double layer resist. A global back gate was provided by the strongly doped silicon Si++ substrate with 285 nm of thermally grown SiO$_2$ on it. Finally, the samples on LOR were current annealed at low temperatures, typically using a bias voltage of $1.6 \pm   0.1$ V. Most of the present work was performed on a Corbino disk with an area of \SI{13}{\micro\metre\squared} and a distance of \SI{1.3}{\micro\metre} between the electrodes (inner and outer radii of 0.9 and \SI{2.25}{\micro\metre}, respectively); a scanning electron microscope picture of a similar Corbino sample is displayed in inset of Fig. \ref{sample}. The gate capacitance $C_g = 1.5 \times 10^{-5}$ F/m$^2$ was obtained using filling factors of Landau levels \cite{Kumar2018}. 

In our experiments, we employed standard voltage-biased measurements for current fluctuations.  The current was amplified using a transimpedance amplifier (SR570, gain $10^5$) and its fluctuations were measured using a Stanford Research SRS 785 FFT analyzer. For details of our experimental techniques we refer to Refs. \onlinecite{Kumar2015a,Laitinen2018d}. 

Kinetic Monte Carlo (kMC) simulations were performed on a model system imitating the Corbino disk geometry by applying appropriate boundary conditions. The calculation procedure and assumptions are described in more detail in Ref. \onlinecite{Kamada2021}. First, the trajectories of the defects on the Corbino disk were generated at two different temperatures ($k_B T = 1.2$ and $k_B T = 2$ while energy barrier for hopping was $E_d=4$) by the kMC simulations allowing 25 defects to move via thermally activated diffusional hops on a 50 by 50 square lattice. The time evolution of the resistance was then calculated by finite element method (FEM) based on the output of the kMC describing the defect motion on the disk. As suggested by experiments with adsorbed Ne \cite{Kamada2021}, the defect sites were assumed to be more conductive than the background lattice. In the present model, the ratio of the conductivity values was set to $10^5$. The FEM calculations were performed for the two orthogonal directions of the current flow, i.e. radial and azimuthal using the same kMC data in both. Finally, a correlation coefficient between the resistance fluctuations in the radial and azimuthal directions was calculated for the two different temperatures.

\end{document}